\documentclass[letterpaper,twocolumn,
prl
,superscriptaddress,email,showpacs
]{revtex4}

\usepackage{soul}
\usepackage[pdftex]{graphicx}%
\usepackage{amssymb}
\usepackage{mathrsfs}
\usepackage{amsmath,amsfonts,latexsym}
\usepackage{array,tabularx,color}
\usepackage{dcolumn} 
\usepackage[normalem]{ulem}
\usepackage{comment}
\usepackage{bm}

\usepackage{doi}
\usepackage{hyperref}

\hypersetup{
 colorlinks=true,
 linkcolor=blue,
 citecolor=red,
 filecolor=red,
 urlcolor=blue,
}

\begin{document}

\title{Cavity-Enhanced Two-Photon Interference using Remote Quantum Dot Sources}

\author{V. Giesz}
\affiliation{Laboratoire de Photonique et de Nanostructures, CNRS, UPR20, Route de Nozay, 91460 Marcoussis, France}

\author{S. L. Portalupi}
\affiliation{Laboratoire de Photonique et de Nanostructures, CNRS, UPR20, Route de Nozay, 91460 Marcoussis, France}
\affiliation{Institut f\"ur Halbleiteroptik und Funktionelle Grenzfl\"achen, Universit\"at Stuttgart, 70569 Stuttgart, Germany}

\author{T. Grange}
\affiliation{Universit\'e Grenoble Alpes, 38000 Grenoble, France}
\affiliation{ CNRS, Institut N\'eel, "Nanophysique et semiconducteurs" group, 38000 Grenoble, France}

\author{C. Ant\'on}
\affiliation{Laboratoire de Photonique et de Nanostructures, CNRS, UPR20, Route de Nozay, 91460 Marcoussis, France}
\affiliation{Departamento de F\'isica de Materiales, Universidad Aut\'onoma de Madrid, Madrid 28049, Spain}

\author{L. De Santis}
\affiliation{Laboratoire de Photonique et de Nanostructures, CNRS, UPR20, Route de Nozay, 91460 Marcoussis, France}

\author{J. Demory}
\affiliation{Laboratoire de Photonique et de Nanostructures, CNRS, UPR20, Route de Nozay, 91460 Marcoussis, France}

\author{N. Somaschi}
\affiliation{Laboratoire de Photonique et de Nanostructures, CNRS, UPR20, Route de Nozay, 91460 Marcoussis, France}

\author{I. Sagnes}
\affiliation{Laboratoire de Photonique et de Nanostructures, CNRS, UPR20, Route de Nozay, 91460 Marcoussis, France}

\author{A. Lema\^itre}
\affiliation{Laboratoire de Photonique et de Nanostructures, CNRS, UPR20, Route de Nozay, 91460 Marcoussis, France}

\author{L. Lanco}
\affiliation{Laboratoire de Photonique et de Nanostructures, CNRS, UPR20, Route de Nozay, 91460 Marcoussis, France}
\affiliation{Universit\'e Paris Diderot - Paris 7, 75205 Paris CEDEX 13, France}

\author{A. Auffeves}
\affiliation{Universit\'e Grenoble Alpes, 38000 Grenoble, France}
\affiliation{ CNRS, Institut N\'eel, "Nanophysique et semiconducteurs" group, 38000 Grenoble, France}

\author{P. Senellart}
\email{pascale.senellart@lpn.cnrs.fr}
\affiliation{Laboratoire de Photonique et de Nanostructures, CNRS, UPR20, Route de Nozay, 91460 Marcoussis, France}
\affiliation{Physics Department, Ecole Polytechnique, F-91128 Palaiseau Cedex, France}

\date{\today}

\begin{abstract}

Quantum dots in cavities have been shown to be very bright  sources of indistinguishable single photons. Yet the quantum interference between two bright quantum dot sources, a critical step for photon based quantum computation, has never been investigated. Here we report on such a measurement, taking advantage of a deterministic fabrication of the devices. We show that cavity quantum electrodynamics can efficiently improve the quantum interference between remote quantum dot sources: poorly indistinguishable photons can still interfere with good contrast with high quality photons emitted by a source in the strong Purcell regime. Our measurements and calculations show that cavity quantum electrodynamics is a powerful tool for interconnecting several devices.

\end{abstract}


\maketitle

Recent years have seen impressive progresses in the implementation of quantum functionalities using semiconductor quantum dots (QDs). The strong anharmonicity of the QD energy levels \cite{Bayer2000} has been used to
generate flying quantum bits in the form of single-photons~\cite{Michler2000} or entangled photon pairs~\cite{Akopian2006,Stevenson2006}, to map and optically manipulate the quantum information encoded onto a single stationary
electron \cite{Emary2007,Press2008} or hole spin \cite{Gerardot2008} as well as to implement optical logic gates \cite{toshibaAPL-CNOT,He2013,Gazzano2013}. The potential of QD-based single-photon sources lies in their
deterministic and pure quantum statistics, as opposed to the parametric down-conversion sources currently used in quantum optics. At saturation, the QD emits a single-photon with a probability close to one, with an
evanescent probability of emitting a second photon. To fully benefit from this statistics, novel strategies have been developed to efficiently collect emitted photons from the high refractive index materials such as inserting the QD in a nanowire on a metallic mirror \cite{Claudon2010} or in a micropillar cavity \cite{Gazzano2013a}. 

Since the first demonstration in 2002 \cite{santori2002a}, the coalescence of photons successively emitted by a single QD has been widely investigated. Various strategies have been developed to minimize the environment-induced dephasing (phonons, charge noise). One approach consists in using a strictly resonant excitation to create directly the carriers into the QD state and reduce the time-jitter of the photon emission \cite{Matthiesen2012,Proux2015,He2013,Wei2014,Muller2014a,valia2014}. Another approach consists in using the Purcell effect by inserting the QD in a microcavity. The acceleration of spontaneous emission reduces the effect of dephasing \cite{santori2002a, Varoutsis2005a,Gazzano2013a} and leads to an efficient extraction of photons: ultra-bright sources of highly indistinguishable photons were recently reported using this approach \cite{Gazzano2013a}.

The scalability of a quantum network based on QDs relies on the possibility of interconnecting two QD devices. Pioneer steps have been made in this direction investigating the two-photon interference between single-photons emitted by remote QDs in planar structures \cite{Flagg2010a,Patel2010b,Gold2014,Gao2013}. Under non-resonant excitation, the coalescence probability is limited to 25-40 \% by the dephasing processes on each source \cite{Flagg2010a,Patel2010b,Gold2014}. The use of a resonant, narrow excitation line has led to a higher coalescence probability (up to 80\%) \cite{Gao2013}. Such technique has not yet been combined  with an efficient extraction of the photons. Besides, it relies on an effective filtering of the events where both QD transitions  are precisely at the laser frequency and is not compatible with a high brightness of the devices. In the present work, we demonstrate that the Purcell effect, which allows efficient collection of photons, is also a powerful tool to improve the coalescence probability from remote sources. We study for the first time the quantum interference of remote QD bright sources, each of them consisting of a single InGaAs QD embedded in a microcavity. Accelerating the spontaneous emission of one source is shown to  improve the two source coalescence probability by efficiently overcoming  the effect of pure dephasing of the other source.

\begin{figure}[hbt!]
\begin{center}
\includegraphics[width=0.95\linewidth]{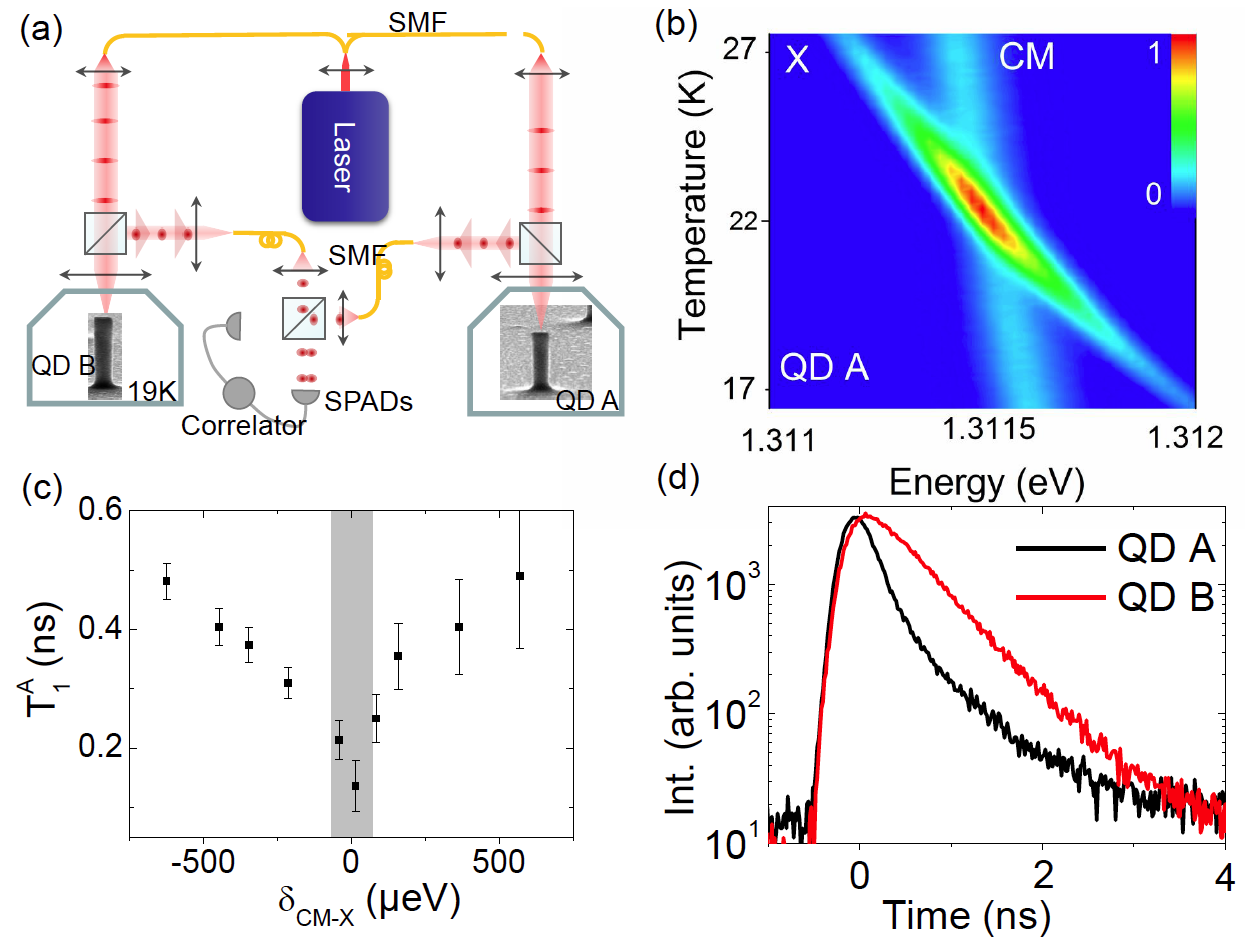}
\caption{(a) Scheme of experimental setup (see text for details). (b) QD A emission as function of temperature and energy. The QD A exciton line (X) is resonant to the bare cavity mode (CM) at 22 K. (c) Decay time of QD A exciton line ($T_1^A$) deconvoluted from the temporal resolution of the setup vs. $\delta_{\mathrm{CM-X}}$. The spectral range used for the two-source interference is marked with a grey, vertical stripe. (d) Decay of the neutral exciton emission of QD A (QD B)  in a black (red) color line for $\delta_{\mathrm{CM-X}}=0$.} \label{fig:1}
\end{center}
\end{figure}

Although it is an essential tool for scalable solid state quantum information processing, remote QD source interference has  been scarcely studied \cite{Flagg2010a,Patel2010b,Gao2013, Gold2014} and never for QDs in cavities which have been shown to be the brightest sources of indistinguishable single photons to date \cite{Gazzano2013a}.  Indeed, this requires to control both the spatial and spectral tuning of each QD to a given cavity mode as well as the respective spectral tuning of the two devices. This highly challenging step is achieved here owing to a deterministic fabrication of the QD-cavity devices.  The micropillar devices are fabricated from the same planar microcavity sample consisting of a bottom (top) Bragg mirror with 32 (18) $\lambda/4$ GaAs/Al$_{0.95}$Ga$_{0.05}$As layers, surrounding an adiabatic cavity embedding a dilute InGaAs QD layer in its center (see Ref.\cite{phonontuned,Lermer2012} for details). Micropillar cavities are laterally centered on single selected QDs with a 50 nm accuracy using the optical \textit{in-situ} lithography technique \cite{Dousse2008}. The cavity diameter is adjusted for each QD so as to match the cavity and QD resonances. Two \textit{in-situ} lithography steps were performed on two parts of the same 1 cm$^2$-area sample so as to fabricate a dozen of micropillar on each. After the pillar etching, the sample was cleaved in two pieces, which were inserted in two separated cryostats with independent temperature-tuning [see Fig. \ref{fig:1}(a)]. Two pillars presenting the same diameter in each cryostat are studied, each of them, deterministically coupled to a QD, referred to as QD A and QD B hereafter.

First, we individually characterize the performances of each single photon source. A single Ti-Sapphire laser, providing 3 ps-long light pulses at a 82 MHz repetition rate, is used to excite both devices, so as to obtain synchronized single-photons from each source. To obtain a good single-photon purity in a regime of strong Purcell effect, the QD devices are excited with a laser energy below the wetting layer resonance \cite{Giesz2013a}. A common excitation state, at 1.370 eV, was found for both QDs.  For both devices, the QD resonance is spectrally very close to the mode resonance at 10 K. The fine spectral matching is obtained by increasing temperatures that shift the exciton lines to lower energies throughout the CM resonance [Fig.\ref{fig:1}(b)]. Figure \ref{fig:1}(c) shows the decay time of the QD A neutral exciton line as a function of the energy detuning $\delta_{\mathrm{CM-X}}= E_{CM} - E_X$. When reducing $\delta_{\mathrm{CM-X}}$, the X decay time decreases, to reach a minimum value of $T_1^A=140\pm40$ ps at resonance ($\delta_{\mathrm{CM-X}}=0$).  A similar study performed on the neutral exciton in QD B gives a lifetime of $T_1^B=470\pm30$ ps at resonance. The difference in lifetime observed for QD A and QD B can be explained by a different coupling to acoustic phonon, resulting is different effective Purcell effect as recently discussed \cite{phonontuned}. The setup used for QD A could be carefully calibrated and  a state of the art brightness was demonstrated: at saturation, a single photon is emitted for each pulse by the QD  and is collected  with a  $74 \pm 5\%$ probability in the first lens. The experimental configuration on the other setup did not allow such a calibration for QD B.

\begin{figure}[h!]
\begin{center}
\includegraphics[width=0.9\linewidth]{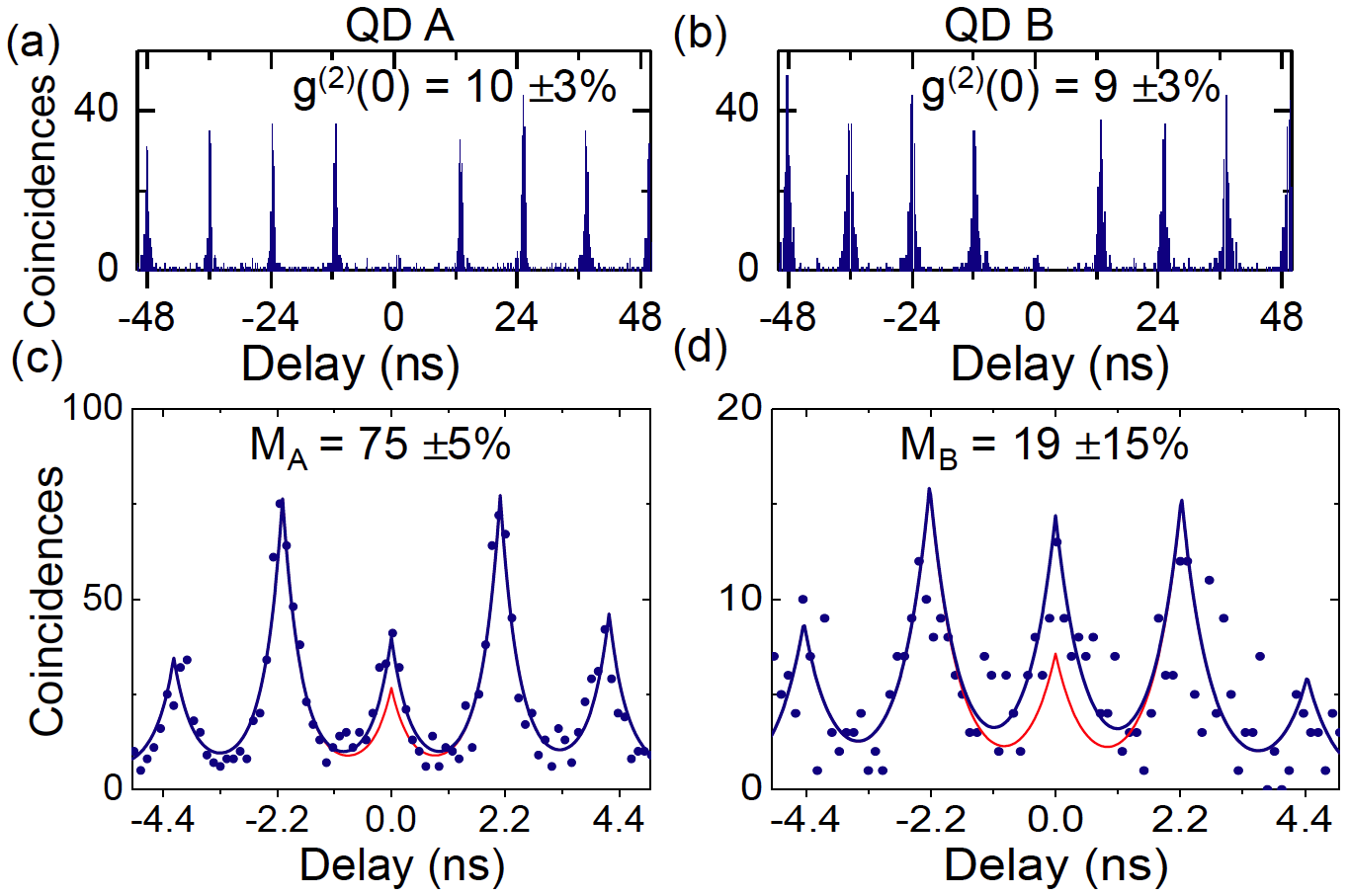}
\caption{Measured autocorrelation function of QD A (a) and QD B (b) exciton lines. (c,d) Correlation histograms measuring the indistinguishability of photons successively emitted by QD A and QD B, respectively. Black lines are a fittings to the experimental data with  $M_A=75 \%$ for QD A and $M_B=19  \%$. As a comparison, the red lines are the calculated curves for $M=1$, considering the measured values for $g^{(2)}(0)$, $\epsilon$, $R$ and $T$ values.} \label{fig:2}
\end{center}
\end{figure}

The single-photon emission is quantified by sending the photons emitted by each device, after coupling into a single mode fiber (SMF), to an a Hanbury Brown-Twiss setup: photons impinge on a 50:50 beam splitter (BS) with outputs coupled to spectrometers for spectral filtering and single-photon avalanche diodes (SPADs).
Figures \ref{fig:2}(a,b) present the autocorrelation functions measured for both devices. A good single-photon purity is obtained for both with an integrated $g^{(2)}(0)$ for the zero delay peaks of $g_A^{(2)}(0)=10 \pm 3\%$ for QD A and $g_B^{(2)}(0)=9 \pm 3\%$ for QD B.

The indistinguishability of successively emitted photons for each source is measured by exciting  each device  twice using a  2.3 ns delay line on the excitation laser line  (see supplementary). The photon indistinguishability is measured through
the mean wave packet overlap ($M$) as defined in Ref. \cite{santori2002a}. For perfectly indistinguishable photons $M=1$, $g^{(2)}(0)=0$ and perfect experimental setup, no signal should be observed at zero delay. Obtaining a good wave packet overlap for successively emitted photons requires choosing properly the excitation wavelength for each source \cite{Gazzano2013a}. With the experimental limitation of using the same excitation line for both sources, we  chose to maximize the indistinguishability on one source (QD A) that also presents the strongest effective Purcell. Under these excitation conditions, we measure  $M$ for each source: it is as high as $M_A=75 \pm 5\%$ for QD A, but only $M_B=19 \pm 15 \%$ for QD B [Figs. \ref{fig:2}(c,d)].

The indistinguishability of photons  emitted by a QD  is mostly limited by charge noise \cite{Gazzano2013a,Gold2014,Kuhlmann2013} that can  lead to a time dependent variation of the X energy (spectral diffusion) or pure dephasing mechanisms, depending on  the time scale of the charge fluctuations  \cite{Berthelot2006a}.   Here, we can equally well  account for all our experimental observations  (including the two-source interference presented below), with both approaches. Thus, we use a pure dephasing description, which allows deriving analytical equations for the two-source interference. In this framework, the  indistinguishability of sequentially emitted photons by a single source is given by $M=\frac{T_2^*}{T_2^*+2T_1}=\frac{\gamma}{\gamma+\gamma^*}$, where $T_1$ ($\gamma$) is the decay time (rate) of the X transition and $T_2^*$ ($\gamma^*/2$) is the pure dephasing time (rate). We deduce  a dephasing $T_2^*$ between $500$ ps and $1500$ ps ($50$ ps and $450$ ps) for QD A (QD B) considering $T_1^A= 140 \pm 40 $ ps ($T_1^B=500  \pm 30$ ps).

\begin{figure}[t]
\begin{center}
\includegraphics[width=.9\linewidth]{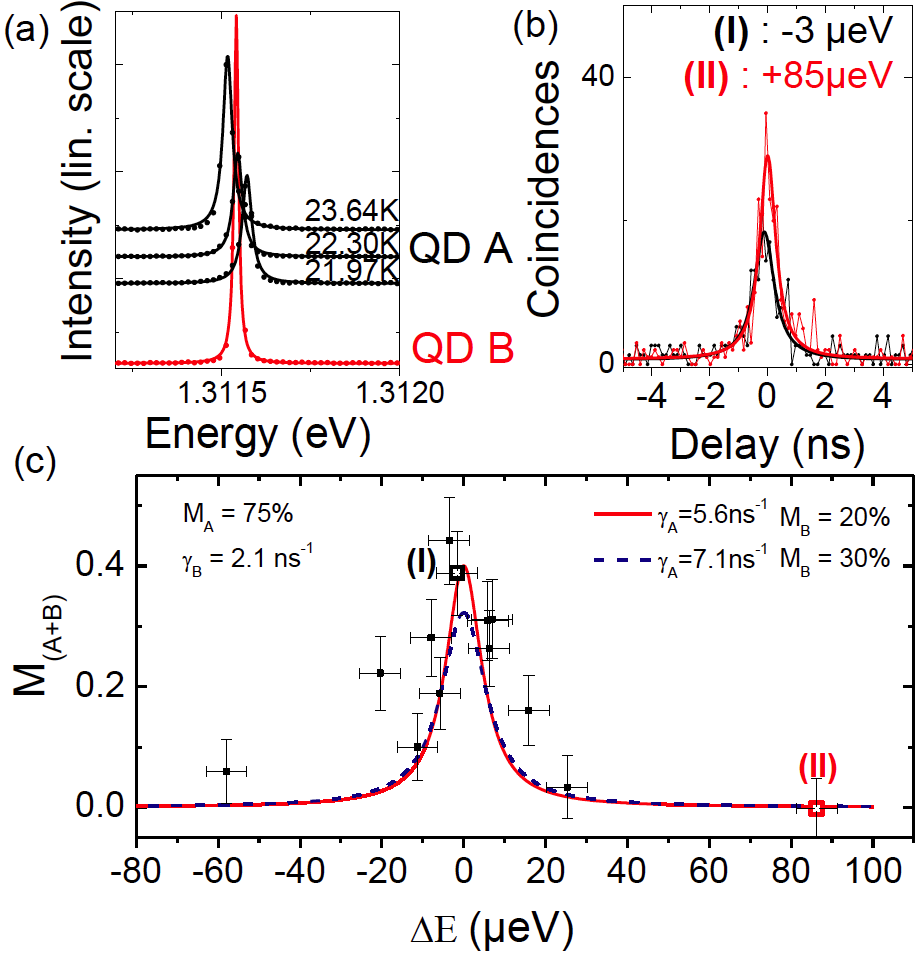}
\caption{(a) Emission spectra for QD B at 19 K (red bottom line) and QD A for various temperatures (black lines) . (b) Measured correlation of the two-source interference at two values of $\Delta E$: (I) $-3$ $\mu$eV (black) and (II)
$+85$ $\mu$eV (red). Full lines are fits to the experimental data. (c) Measured $M_{(A+B)}$ as a function of $\Delta E$. Lines are fits to the experimental data using the parameters indicated in the legend. The particular detunings (I) and (II) are labeled in the panel.} \label{fig:3}
\end{center}
\end{figure}

We now characterize the two-source interference by measuring the photon coalescence  for different spectral detunings of the sources:  QD B is kept to a fixed temperature where $\delta_{\mathrm{CM-X}}^B=0$ and the temperature of QD A is varied over 1.7 K to tune the spectral detuning $\Delta E=E_X^{B}-E_X^{A}$ of the two sources [Fig. \ref{fig:3}(a)]. The QD excitations are respectively delayed so that the generated single-photons impinge simultaneously  on the two inputs of the BS. The photon correlation on the two output SPADs allows deducing the mean wave packet overlap for the two sources $M_{(A+B)}$ using the area $A_0$ of the zero delay peak (see supplementary). Figure \ref{fig:3}(b) presents the zero delay peak of the correlation curve for the sources detuning $\Delta E=- 3 $ $\mu$eV (black) and $+85 $ $\mu$eV (red). A clear decrease of the zero delay peak area is observed when the two sources are in resonance, showing the increased photon coalescence probability. Figure \ref{fig:3}(c) presents the measured $M_{(A+B)}$ as a function of $\Delta E$. For large detunings, it is observed that $M_{(A+B)}=0$, as expected for distinguishable photons. At $\Delta E=0$, a mean wavepacket overlap  $M_{(A+B)}=40\pm4\%$ is observed,  a value largely exceeding $M_B$. Moreover, the two-photon interference takes place on a spectrally broad spectral range, with a full
width at half maximum of $\sim15 \pm 5$ $\mu$eV, three time larger than previous observations with QDs in planar structure \cite{Patel2010b}. In the following, we discuss  how the two source interference is enhanced by use of Purcell effect.

If both sources undergo pure dephasing, $M_{(A+B)}$ for the two photons overlap is:
\begin{equation}
M_{(A+B)}=\frac{\gamma_A \gamma_B}{\gamma_A+\gamma_B} \: \frac{\gamma_A+\gamma_B+\gamma_A^*+\gamma_B^*}{(\frac{\Delta E}{\hbar})^2 + \frac{1}{4}(\gamma_A+\gamma_B+\gamma_A^*+\gamma_B^*)^2}
\label{eq:eq3}
\end{equation}
(see supplementary)

 The linewidth of the two photon interference is thus given by the sum of each emitter linewidth. The strong Purcell effect on QD A, contributing through $\gamma_A$, leads to a significant increase of the spectral range over which the two-source quantum interference takes place. 
 
 As shown in Fig. \ref{fig:3}(c), our experimental observations can well be reproduced with $M_A$=0.75,  $\gamma_B=2.1$ ns$^{-1}$, $5.6 < \gamma_A < 7.1$ ns$^{-1}$  and $0.2<M_B<0.3$, which are consistent with the measurements performed on each source within the error bars.
 The  two-source coalescence probability reaches $M_{A+B}=40\pm 4 \%$, an intermediate value between $M_A=75 \pm 5\%$ and $M_B=19\pm15\%$. Yet, as discussed now, this value does not result from  a simple averaging of the values obtained for each source. Controlling the spontaneous emission on QD A actually significantly improves $M_{A+B}.$

\begin{figure}[htb!]
\begin{center}
\includegraphics[width=0.8\linewidth]{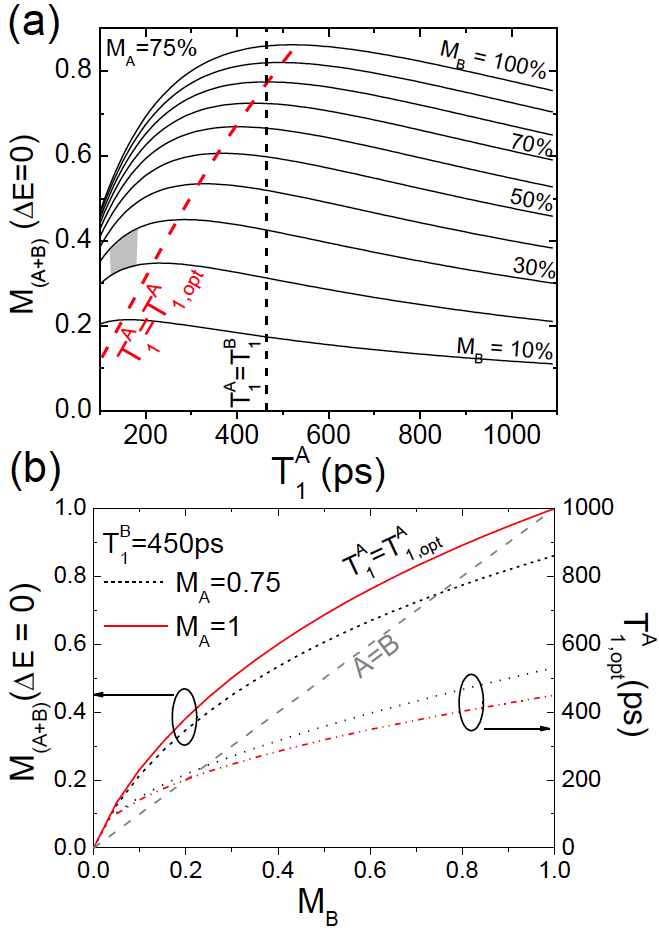}
\caption{(a) Calculated photon overlap $M_{(A+B)}$ at $\Delta E = 0$  as a function of QD A decay time $T_1^A$ (with a fixed $M_A=75\%$) for different values of $M_B$. The decay time of QD B is fixed at $T_{1}^B=470$ ps. Red, dashed line indicates the optimal values of $T_{1,opt}^A$ where $M_{(A+B)}(\Delta E = 0)$ is maximum. The vertical, dashed line indicates the situation where the two sources present the same decay time. (b) (Left, vertical axis) Optimal photon overlap $M_{(A+B)}(\Delta E = 0)$ obtained at $T_{1,A}=T_{1,opt}^{A}$ and $T_{1}^B=470$ ps in a full, red (dashed, black) line for $M_A=1$ ($M_A=0.75$). The grey, dashed line indicates the indistinguishability when both sources are identical. (Right, vertical axis) $T_{1,opt}^A$ decay time as function of $M_B$ for a value of $M_A=1$ (0.75) in a red, dot-dashed (black, dotted) line.} \label{fig:4}
\end{center}
\end{figure}

Figure \ref{fig:4}(a) presents the calculated $M_{(A+B)}$ for $\Delta E = 0$ as a function of the decay time of QD A for various values of $M_B$. The lifetime of QD B is fixed to $T_{1}^B= 450$ ps and $M_A=75 \%$. The
vertical, dashed line indicates the situation where the two sources present the same decay time. The coalescence probability is optimal for sources with identical lifetime only for $M_A\approx M_B$. When $M_B < M_A$, the
maximum probability of coalescence is obtained for an optimal value of $T_{1}^A$ ($T_{1,opt}^A$) that depends on $M_B$: the lower the indistinguishability of successively emitted photons by QD B source, the stronger the
Purcell factor on QD A should be to overcome the pure dephasing on QD B. In other words, the faster the decoherent processes on $X_B$ are, the faster QD A emission should be to compensate them. The grey area in Fig. \ref{fig:4}(a) presents the conditions reached in the current measurements where this effect is demonstrated.

%
%

For the optimal value $T_{1,opt}^{A}$, the coalescence probability of the two-source interference significantly exceeds $M_B$. Figure \ref{fig:4}(b) presents in the left, vertical axis the best achievable overlap $M_{(A+B)}$ as a function of $M_B$ obtained with $T_{1}^A=T_{1,opt}^{A}$ for the case of the present measurement $M_A=75\%$ (black, dotted line) and for $M_A=100\%$ (red, full line). The grey, dashed line allows comparing the best achievable overlap to the case where both sources are identical. The right, vertical axis shows the value of $T_{1,opt}^A$ as function of $M_B$ that renders the optimal coalescence probability
$M_{(A+B)}$. For $M_B<M_A$, the best achievable overlap in the two-source experiment is significantly larger than the one obtained for photons successively emitted by QD B and exceeds the one obtained considering a source with the same lifetime.
 
In solid-state systems, where dephasing is a limitation to scalability, cavity quantum electrodynamics is shown, for the first time in a two-source experiment, to be a powerful asset. The Purcell effect, known to improve the indistinguishability of photons successively emitted by a single source, is shown to enhance the quantum interference of remote sources. This result is crucial for the scalability of QD-based quantum networks, where the imperfections of one device can be efficiently compensated by a highly indistinguishable single-photon source with a controlled lifetime. 

\begin{acknowledgments}
This work was partially supported by the ERC starting grant 277885 QD-CQED, the French RENATECH network, the Labex NanoSaclay, the CHISTERA project SSQN and the EU FP7 grant 618072 (WASPS). C.A. acknowledges financial support from the Spanish FPU scholarship.
\end{acknowledgments}

%


\widetext
\clearpage
\begin{center}
\textbf{\large Supplemental Materials: Cavity-Enhanced Two-Photon Interference using Remote Quantum Dot Sources}
\end{center}
\setcounter{equation}{0}
\setcounter{figure}{0}
\setcounter{table}{0}
\setcounter{page}{1}
\makeatletter
\renewcommand{\theequation}{S\arabic{equation}}
\renewcommand{\thefigure}{S\arabic{figure}}
\renewcommand{\bibnumfmt}[1]{[S#1]}
\renewcommand{\citenumfont}[1]{S#1}

\begin{center}
V. Giesz,$^1$ S. L. Portalupi,$^{1, 2}$ T. Grange,$^{3, 4}$ C. Ant\'on,$^{1, 5}$ L. De Santis,$^{1}$ J. Demory,$^{1}$ N. Somaschi,$^{1}$ I. Sagnes,$^{1}$ A. Lema\^itre,$^{1}$ L. Lanco,$^{1, 6}$ A. Auffeves,$^{3, 4}$ and P. Senellart$^{1, 7, *}$
\end{center}

\begin{center}
\textit{$^1$Laboratoire de Photonique et de Nanostructures, CNRS, UPR20, Route de Nozay, 91460 Marcoussis, France \\
$^2$Institut f\"ur Halbleiteroptik und Funktionelle Grenzfl\"achen, Universit\"at Stuttgart, 70569 Stuttgart, Germany \\
$^3$Universit\'e Grenoble Alpes, 38000 Grenoble, France \\
$^4$CNRS, Institut N\'eel, ``Nanophysique et semiconducteurs" group, 38000 Grenoble, France  \\
$^5$Departamento de F\'isica de Materiales, Universidad Aut\'onoma de Madrid, Madrid 28049, Spain \\
$^6$Universit\'e Paris Diderot - Paris 7, 75205 Paris CEDEX 13, France \\
$^7$Physics Department, Ecole Polytechnique, F-91128 Palaiseau Cedex, France}
\end{center}

\section{Experimental methods}
\subsection{Indistinguishability of successively emitted photons}
To measure the indistinguishability of successively emitted photons for each source, each device is excited twice using an additional 2.3 ns delay line on the excitation laser line. The emitted single-photons, coupled into the SMF, are non-deterministically split using a 50:50
fiber-based BS and then sent to the two inputs of a cube BS. A 2.3 ns fiber delay line is added on one of the outputs of the fiber BS so that the two photons reach the BS at the same time. A polarizer is inserted between the collection objective and the SMF to collect photons polarized along the QD axes. This polarization is further adjusted before the two inputs of the BSs with polarizers and fiber-paddles. The temporal coincidences are measured with two SPADs at the outputs of the cube BS after spectral filtering with spectrometers (17 pm resolution). The photon indistinguishability is measured through
the mean wave packet overlap ($M$) as defined in Ref. \cite{santori2002a}.  For the present experimental configuration, the measured area of the peak at zero delay ($A_0$) compared to the area of the lateral peaks at $\pm$ 2.3 ns ($A_{\pm 2.3\mathrm{ns}}$) allows deducing $M$ using:

\small
\begin{align}
M= \frac{1}{(1-\epsilon)^2} \Big[& 2g^{(2)}(0)+\frac{R^2+T^2}{2RT} \nonumber \\ &- \frac{A_0}{A_{-2.3\mathrm{ns}}+A_{+2.3\mathrm{ns}}} \left( 2+g^{(2)}(0)\frac{(R^2+T^2)}{RT} \right) \Big]
\label{eq:eq1}
\end{align}
\normalsize

\noindent where $\epsilon=0.88\pm0.03$ is the interference fringe contrast measured with a coherent laser, and $R=0.40\pm0.02$ and $T=0.50\pm0.02$ are the BS intensity reflection and transmission, respectively.  

\subsection{Coalescence with remote sources}

To measure the indistinguishability of photon emitted by QD A and QD B, the QD excitations are respectively delayed so that the generated single-photons impinge simultaneously (with a temporal accuracy of $\sim8$ ps) on the two inputs of the BS. The photon correlation on the two output SPADs allows deducing the mean wave packet overlap using the area $A_0$ of the zero delay peak and the area $A_{12.2 \mathrm{ns}}$ of the 12.2 ns peaks coming from the laser repetition rate:

\small
\begin{align}
M_{(A+B)} = \frac{1}{\left(1-\epsilon\right)^2} \Bigg[&\frac{{g^{(2)}}_A(0)+{g^{(2)}}_B(0)}{2} + \frac{R^2+T^2}{2RT} \nonumber \\ &- \frac{(R+T)^2}{2RT} \frac{A_0}{A_{12.2 \mathrm{ns}}} \Bigg]
\label{eq:eq2}
\end{align}
\normalsize

\section{Theory of the two-source wave packet overlap}

We consider two photon wave packets sent on the two input ports (denoted A,B) of a beam splitter. The normalized probability $p_{\text{c}}$ of coincidence detection in the two output ports (denoted C,D) reads \cite{kiraz2004quantum}
\begin{equation}
p_{c} = \frac{\int dt_1 \int dt_2  g^{(2)}_{CD}(t_1,t_2) }{ \left[\int dt \langle \hat{a}_C^{\dagger}(t) \hat{a}_C(t) \rangle \right]\left[\int dt \langle \hat{a}_D^{\dagger}(t) \hat{a}_D(t) \rangle \right]},
\label{pc}
\end{equation}
with
\begin{equation}
g^{(2)}_{CD}(t_1,t_2) = \langle \hat{a}_C^{\dagger}(t_1) \hat{a}_D^{\dagger}(t_2) \hat{a}_D(t_2) \hat{a}_C(t_1)  \rangle.
\end{equation}

The beam splitter generates a unitary transformation of the form 
\begin{equation}
\begin{pmatrix}
\hat{a}_C \\ \hat{a}_D
\end{pmatrix}
=
\begin{pmatrix}
\sqrt{T} & - \sqrt{R} e^{-i\phi}\\
\sqrt{R} e^{i\phi} & \sqrt{T} 
\end{pmatrix}
\begin{pmatrix}
\hat{a}_A \\ \hat{a}_B
\end{pmatrix}.
\end{equation}

In the limit of single photon wavepackets (i.e. $g^{(2)}_{AA}=0$ and $g^{(2)}_{BB}=0$), and assuming no phase correlations between the two sources, it can be expressed in terms of the input fields A and B as:
\begin{equation}
\begin{split}
g^{(2)}_{CD}(t,t+\tau) = T^2
\langle a_A^{\dagger}(t)a_A(t)\rangle \langle a^{\dagger}_B(t+\tau)a_B(t+\tau) \rangle 
+ R^2 \langle a^{\dagger}_A(t+\tau)a_A(t+\tau) \rangle \langle a_B^{\dagger}(t)a_B(t)\rangle \\
- TR \langle a_A^{\dagger}(t)a_A(t+\tau)\rangle \langle a^{\dagger}_B(t+\tau)a_B(t) \rangle 
- TR \langle a_A^{\dagger}(t+\tau)a_A(t)\rangle \langle a^{\dagger}_B(t)a_B(t+\tau) \rangle. \\
\end{split}
\end{equation}
Within the Markov approximation, the wave packets sent by the two-level systems in A and B initially excited at $t=0$ reads:

\begin{equation}
\langle a_A^{\dagger}(t)a_A(t+\tau)\rangle = n_A \gamma_A e^{-\gamma_A t} e^{-i\omega_A \tau} e^{-(\gamma_A+\gamma^*_A) \tau/2},
\end{equation}

\begin{equation}
\langle a_B^{\dagger}(t)a_B(t+\tau)\rangle = n_B \gamma_B e^{-\gamma_B t} e^{-i\omega_B \tau}
e^{-(\gamma_B+\gamma^*_B) \tau/2} ,
\end{equation}
where $\gamma_A$ ($\gamma_B$) is the radiative decay rate of the emitter A (resp. B), $\omega_A$ ($\omega_B$) is its angular frequency, $\gamma_A^*/2$ ($\gamma_B^*/2$) is its pure dephasing rate, and  $n_A$ ($n_B$) is the probability of photon emission per wave packet for the emitter A (resp. B).
Hence $g^{(2)}_{CD}$ can be rewritten as
\begin{equation}
\begin{split}
g^{(2)}_{CD}(t,t+\tau) = n_A n_B T^2
e^{-(\gamma_A+\gamma_B) t}e^{-\gamma_B \tau}
+  n_A n_B  R^2 e^{-(\gamma_A+\gamma_B) t}e^{-\gamma_A \tau} \\
-  n_A n_B TR e^{-(\gamma_A+\gamma_B) t}e^{-(\gamma_A+\gamma^*_A+\gamma_B+\gamma^*_B) \tau/2}\left[ e^{i\Delta \omega \tau}+e^{-i\Delta \omega \tau} \right]. \\
\end{split}
\end{equation}
where $\Delta \omega = \omega_B-\omega_A$.
Integrating over the two time variables gives
\begin{equation}
\begin{split}
\int_0^{\infty} dt_1 \int_0^{\infty}  d_2 t_2 ~ g^{(2)}_{CD}(t_1,t_2) & =
2\int_0^{\infty} dt \int_0^{\infty}  d \tau ~ g^{(2)}_{CD}(t,t+\tau) \\
 & = \frac{2n_A n_B}{\gamma_A+\gamma_B} \int_0^{\infty} d\tau \left[T^2 e^{-\gamma_B \tau}
+  R^2 e^{-\gamma_A \tau}-   TR e^{-(\gamma_A+\gamma^*_A+\gamma_B+\gamma^*_B) \tau/2}\left( e^{i\Delta \omega \tau}+e^{-i\Delta \omega \tau} \right) \right]   \\
 & =  \frac{2n_A n_B}{\gamma_A+\gamma_B} \left[ \frac{T^2}{\gamma_B} + \frac{R^2}{\gamma_A} 
-TR \frac{(\gamma_A+\gamma_B+\gamma^*_A+\gamma^*_B)}{\Delta \omega^2 + \left[ (\gamma_A+\gamma_B+\gamma^*_A+\gamma^*_B)/2 \right]^2}
\right].
\end{split}
\label{intgCD}
\end{equation}
On the other hand we have
\begin{equation}
\begin{split}
\int_0^{\infty} dt \langle \hat{a}_C^{\dagger}(t) \hat{a}_C(t) \rangle 
& =  \int_0^{\infty} dt \left[ T \langle \hat{a}_A^{\dagger}(t) \hat{a}_A(t) \rangle + R\langle \hat{a}_B^{\dagger}(t) \hat{a}_B(t) \rangle \right]\\
& = \frac{Tn_A}{\gamma_A}+\frac{Rn_B}{\gamma_B} , \\
\end{split}
\label{intC}
\end{equation}
\begin{equation}
\begin{split}
\int_0^{\infty} dt \langle \hat{a}_D^{\dagger}(t) \hat{a}_D(t) \rangle 
& = \frac{Rn_A}{\gamma_A}+\frac{Tn_B}{\gamma_B} .\\
\end{split}
\label{intD}
\end{equation}
For a balanced beam splitter (i.e. $T=R=1/2$), by plugging Eqs.~\ref{intgCD},  \ref{intC} and  \ref{intD} in Eq.~\ref{pc} we obtain
\begin{equation}
\begin{split}
p_c = \frac{1}{2} (1 
- M_{(A+B)} )
\end{split}
\end{equation}
where $M_{(A+B)}$, defined as the mean wave packet overlap, is given by 
\begin{equation}
\begin{split}
M_{(A+B)}= \frac{\gamma_A \gamma_B}{\gamma_A + \gamma_B} \frac{(\gamma_A+\gamma_B+\gamma^*_A+\gamma^*_B)}{\Delta \omega^2 + \left[ (\gamma_A+\gamma_B+\gamma^*_A+\gamma^*_B)/2 \right]^2}.
\end{split}
\end{equation}

\end{document}